\documentclass[epj,nopacs]{svjour}
\renewcommand{\bar}[1]{\overline{#1}}

\usepackage{graphics}
\begin{document}
\title{Neutrons and antiprotons in ultrahigh energy cosmic rays}

\author{W-Y. P. Hwang \inst{1} \and Bo-Qiang Ma\inst{2,}\thanks{Corresponding author. Mailing Address:
Department of Physics, Peking University, Beijing 100871, China. e-mail: mabq@phy.pku.edu.cn }}

%
%
\institute{Department of Physics, National Taiwan University,
Taipei, Taiwan 10617 \and Department of Physics, Peking
University, Beijing 100871}

\date{Received: date / Revised version: date}
%
\abstract{The neutron fraction in the very high energy cosmic rays
near the Greisen-Zatsepin-Kuzmin (GZK) cutoff energy is analyzed
by taking into account the time dilation effect of the neutron
decays and also the pion photoproduction behaviors above the GZK
cutoff. We predict a non-trivial neutron fraction above the GZK
cutoff and a negligibly small neutron fraction below. However,
there should be a large antiproton fraction in the high energy
cosmic rays below the GZK cutoff in several existing models for
the observed cosmic-ray events above and near the GZK cutoff. Such
a large antiproton fraction can manifest itself by the muon charge
ratio $\mu^+/\mu^-$ in the collisions of the primary nucleon
cosmic rays with the atmosphere, if there is no neutron
contribution. We suggest to use the muon charge ratio as one of
the information to detect the composition of the primary cosmic
rays near or below the GZK cutoff.
} 

\titlerunning{Neutrons and antiprotons in ultrahigh energy cosmic rays}
\maketitle


The observation of ultrahigh energy cosmic rays above and near the
Greisen-Zatsepin-Kuzmin (GZK) cutoff energy presents an
outstanding puzzle in astrophysics and cosmology \cite{review}. It
has long been anticipated that the highest energy cosmic rays
would be protons from outside the galaxy, and there is an upper
limit of the highest energy in the observed proton spectrum,
commonly referred to as the GZK cutoff \cite{GZK}, as the protons
travelling from intergalactic distances should experience energy
losses owing to pion productions by the photons in the cosmic
background radiation. Then it is suggested \cite{Stecker} 
that the particles with energy above the GZK cutoff must be coming
from within the local ``supercluster" of galaxies of which we are
a part. Thus, the ``GZK cutoff" is not a true cutoff, but a
suppression of the ultrahigh energy cosmic ray flux owing to a
limitation of the propagation distance, which we refer to as the
GZK zone. Although there have been some novel explanations for the
observed cosmic ray events above the GZK cutoff, it is natural to
expect that these ultrahigh energy cosmic rays come from sources
within the GZK zone, i.e., not far from us in more than tens of
Mpc. The ``Z-bursts" \cite{Weiler,FMS,GK} can
explain the highest energy cosmic ray events, 
with sources within the GZK zone \cite{Stecker}. The reason is
that the ``Z-bursts" are from the $Z$-boson annihilations of the
ultrahigh energy neutrino (antineutrino) cosmic rays with the
relic neutrinos (antineutrinos) in the cosmic background
radiation, and these annihilations can happen anywhere in the
universe. The ``Z-bursts" can produce nucleon cosmic rays with
ultrahigh energy within the GZK zone, as the energy of the
produced Z-bosons is high enough by the collision of ultrahigh
energy neutrino beams with the relic neutrinos of non-zero mass.
Therefore the nucleons with ultrahigh energy from ``Z-bursts" can
reach us without the GZK constraint, and this can explain the
observed cosmic ray events with energy above and near the GZK
cutoff. Several existing models of similar kinds, such as using
relic metastable superheavy particles \cite{X} and topological
defects \cite{Bha98} as origins of ultrahigh energy particles, can
also explain the ultrahigh energy cosmic ray events above and near
the GZK cutoff.

In fact, these models can produce a number of photons and nucleons
to reach the earth, for example, the photon/nucleon ratio is of
about 10 on average in the Z-burst model. A possible method to
test these models is by the large $\gamma/p$ ratio for the cosmic
rays near the upper end of the spectrum. This can be achieved by
the separation of primary photon and proton cosmic ray events with
different characteristic air shower profiles. However, for the
nucleon cosmic rays from these models, there could be a sizable
number of neutrons and antinucleons in comparison with that of the
protons. For example, there are roughly equally numbers of protons
and neutrons at the source point in the Z-burst model, with also a
symmetric part of antinucleons. Therefore it is necessary to
estimate the composition of the nucleon cosmic rays. We will study
the role played by the neutrons in the ultrahigh energy cosmic
rays near the GZK cutoff. We will take into account the the time
dilation effect of the neutron decays and the pion photoproduction
behaviors of the nucleons above the GZK cutoff. It will be shown
that there should be a non-trivial fraction of neutrons above the
GZK cutoff, whereas the neutron fraction is negligibly small below
the GZK cutoff. Since there should be equally numbers of protons
and antiprotons in the cosmic rays from the Z-bursts, and these
protons and antiprotons will produce different muon charge ratio
$\mu^+/\mu^-$ in the air showers, we thus suggest to measure the
$\mu^+/\mu^-$ ratio of the air showers by the primary cosmic rays
near or below the GZK cutoff, as one of the information to detect
the composition of the primary cosmic rays with high energy.

The 2.73 K cosmic microwave background (CMB) of the photons
satisfies the Planck's ideal black-body radiation formula, with
the number density $n_{\gamma}=16\pi \zeta (3) \left(kT/hc
\right)^3=413$ photons per cm$^3$ and the mean energy per photon
$\epsilon_{\gamma}=\pi^4kT/30 \zeta (3)=6.35 \times 10^{-4}$~eV,
where $\zeta (3)=1.20$ is the Riemann Zeta function. When the
nucleon with 4-momentum $p=(E,{\mathbf p})$ interacts with the
photon with 4-momentum $k=(\epsilon, {\mathbf k})$, and composes
into a system with the center of mass energy squared $S$, we have
\begin{equation}
E=\left(S-m_N^2\right)/2\epsilon \left(1-\sqrt{1-m_N^2/E^2}\cos
\theta\right),
\end{equation}
where $\theta$ is the angle between ${\mathbf p}$ and ${\mathbf
k}$. $\theta$ cannot be zero since a nucleon cannot catch up a
photon moving in the same direction, and the energy of the nucleon
$E$ must be very large near the pion photoproduction process
$N+\gamma_{CMB}\to \pi +N$, therefore we have,
\begin{equation}
E \approx \left(S-m_N^2\right)/2\epsilon \left(1- \cos \theta
\right).
\end{equation}
The threshold energy for pion production $N+\gamma_{CMB}\to \pi
+N$ is
\begin{equation}
E \approx \left(2m_N m_{\pi}+m_{\pi}^2\right)/4\epsilon =1.10
\times 10^{20}~ {\mathrm{eV}} ,
\end{equation}
and the threshold energy for producing the $\Delta$ resonance
$N+\gamma_{CMB} \to \Delta \to \pi +N$ is
\begin{equation}
E \approx \left(m_{\Delta}^2-m_{N}^2\right)/4\epsilon =2.52 \times
10^{20}~ {\mathrm{eV}} .
\end{equation}
The neutron has a mean life time $\tau_{n}=887$ s in its rest
reference frame. Due to the time dilation effect of the special
relativity, the life time of a moving particle is dilated by a
factor $\gamma_{n}=E_n/m_n$. Therefore the mean free path that a
moving neutron can travel before its beta decay into the proton
should be, for a neutron at the pion production threshold,
\begin{equation}
l_n \approx c \gamma_{n} \tau_n =3.12 \times
10^{24}~{\mathrm{cm}}=1.01~ {\mathrm{Mpc}},
\end{equation}
and for a neutron at the $\Delta$ resonance threshold,
\begin{equation}
l_n \approx c \gamma_{n} \tau_n =7.11 \times
10^{24}~{\mathrm{cm}}=2.30~ {\mathrm{Mpc}}.
\end{equation}
Therefore the neutron fraction around the GZK cutoff is expected
to be negligibly small for nucleons coming from a source with
distance of considerably more than a few Mpc, if we only consider
the time dilation effect of the neutrons.

However, the situation is different if we take into account the
detailed features for the pion photoproduction $N +\gamma_{CMB}
\to \pi +N$, which, when distinguishing between proton and
neutron, should be written as $p+\gamma_{CMB} \to \pi^+ +n$,
$n+\gamma_{CMB} \to \pi^- +p$, $p+\gamma_{CMB} \to \pi^0 +p$, and
$n+\gamma_{CMB} \to \pi^0 + n$.
These four processes have quite different cross sections, as can
be understood from the low energy theorem \cite{LET} as well as
from the chiral quark model \cite{CQM}. Roughly speaking, the
cross section of the pion photoproduction $N(\gamma,\pi)N$ is
proportional to $A^2$ with the four physical amplitudes of $A$
expressed by \cite{CQM}
\begin{eqnarray}
A(p \gamma \to \pi^+ n)&=&\sqrt{2}\left(A^{-}+A^{0}\right),\\
A(n \gamma \to \pi^- p)&=&\sqrt{2}\left(A^{-}-A^{0}\right),\\
A(p \gamma \to \pi^0 p)&=&A^{+}+A^{0},\\
A(n \gamma \to \pi^0 n)&=&A^{+}-A^{0},
\end{eqnarray}
where the isospin amplitudes can be expanded in terms of the ratio
between pion mass and nucleon mass $\eta=m_{\pi}/m_N$
\begin{equation}
A^{-}=1+O(\eta^2), ~~ A^{+}=A^0=-\eta/2+O(\eta^2).
\end{equation}
Thus we get
\begin{equation}
\frac{\sigma(n \gamma  \to \pi^- p)}{\sigma(p \gamma \to \pi^+
n)}=\frac{(1+\eta/2)^2}{(1-\eta/2)^2} \approx 1.34,
\end{equation}
which is in excellent agreement with the experimental data
\cite{piondata}, and
\begin{eqnarray}
\frac{\sigma(p \gamma  \to \pi^0 p)}{\sigma(p \gamma \to \pi^+
n)}&=&\frac{\eta^2}{2(1-\eta/2)^2} \approx 0.01,\\
\frac{\sigma(n \gamma  \to \pi^0 n)}{\sigma(p \gamma \to \pi^+
n)}&=&\frac{O(\eta^4)}{2(1-\eta/2)^2} \approx O(\eta^4),
\end{eqnarray}
which means that the neutral pion production processes,
$p+\gamma_{CMB} \to \pi^0 +p$ and $n+\gamma_{CMB} \to \pi^0 + n$,
can be neglected. Adopting an average cross section $\sigma(p
\gamma \to \pi^+ n)=200$~$\mu{\mathrm{b}}$ \cite{GZK,Stecker}
above the pion photoproduction threshold, we have the mean free
path of interaction for the proton
\begin{equation}
\lambda_p=\frac{1}{n_{\gamma}\sigma(p \gamma \to \pi^+ n)}
=1.21\times 10^{25}~{\mathrm{cm}}=3.92~{\mathrm{Mpc}},
\end{equation}
and that for the neutron
\begin{equation}
\lambda_n=\frac{1}{n_{\gamma}\sigma(n \gamma \to \pi^- p)}
=9.04\times 10^{24}~{\mathrm{cm}}=2.93~{\mathrm{Mpc}}.
\end{equation}

It is interesting to notice that the protons and neutrons change
into each other via charged pion production by the relic photons
in the travel until the nucleon energies degraded to below the GZK
cutoff. There is always a certain amount of neutron fraction for
nucleons above the pion photoproduction threshold, since the
protons can always change into neutrons via the charged pion
photoproduction. Though the neutrons change more fast into protons
via both beta decay or charged pion photoproduction (with
effective mean free path
$\lambda_n^{\mathrm{eff}}=l_n\lambda_n/(l_n+\lambda_n)$), these
produced protons continue to change into neutrons if their energy
is still above the pion production threshold. As a consequence,
there is always a non-trivial neutron fraction in the nucleon
cosmic rays above the GZK cutoff. The magnitude of the neutron
fraction depends on the energy of the nucleon and the sources, and
it increases with the increase of energy, as the time dilation
effect is stronger for neutrons with higher energy. Detailed
features of the neutron fraction and nucleon spectrum may also
provide information on the sources of the nucleon cosmic rays,
such as their distance, the nucleon spectrum of the initial
sources and their neutron/proton ratios. For example, assuming
that the nucleons with energy above the GZK cutoff are from a
point source with an uniform nucleon spectrum (i.e., with a
constant value of density distribution) at a distance far away
(explicit calculation shows that $3 \lambda_p$ is enough), the
neutron/proton ratio should reach the equilibrium value of
$\lambda_n^{\mathrm{eff}}/\lambda_p \approx 0.19$ at the pion
production threshold and $0.33$ at the $\Delta$ production
threshold, independent of the neutron/proton ratio of the source.
Adopting a small cross section at threshold could reduce
$\lambda_n^{\mathrm{eff}}/\lambda_p $ to around 0.05, in agreement
with explicit model calculation~\cite{HLM}. Thus, there must be a
significant neutron fraction in the nucleon cosmic rays above the
GZK cutoff, even if the sources only emit protons.

However, the neutron fraction is negligibly small below the GZK
cutoff, since only the neutrons within a distance comparable to
the mean free path of decay $l_n \approx 1$ Mpc can reach the
earth. The protons with energy below the GZK cutoff may interact
with the relic photons via electron-position pair production, but
the energy loss is small, of only 0.1\% compared to 20\% for pion
production. Therefore such protons may come from any source with
distance within as large as half the Hubble length \cite{GZK}. We
thus expect a small neutron/proton ratio below the GZK cutoff.

For the Z-bursts of particle productions, there should be a
symmetry between the nucleon and antinucleon productions in the
standard model. Thus there should be equally numbers of protons
and antiprotons in the nucleon cosmic rays from the Z-bursts.
Balloon and satellite investigations can provide direct
measurements of the cosmic rays of proton and antiprotons
respectively. Unfortunately, the available measurements can only
reach to the energy scale of $10^{15}$~eV \cite{Wef03}, still $4
\to 5$ orders below the energy region for our purpose. One of the
methods to identify the hadron species in the cosmic rays is by
measuring the muon charge ratio $\mu^+/\mu^-$ of the air shower by
the primary cosmic rays. It has been known that the $\mu^+/\mu^-$
ratio may provide information on the neutron/proton ratio in the
primary cosmic rays \cite{Adair}. With the improvements in
knowledge of particle productions in hadron-hadron interactions
from accelerator experiments and also in understanding of particle
production and propagation mechanism in the atmosphere, the
$\mu^+/\mu^-$ ratio, in combination with other coincidence
measurements, can reveal the species in the primary beams.
However, we will show that the $\mu^+/\mu^-$ ratio should be
similar for primary neutron and antiproton beams. Thus the
$\mu^+/\mu^-$ ratio is hard to distinguish between the neutron and
antiproton in the nucleon cosmic rays. With the above prediction
that the neutron fraction is negligibly small below the GZK
cutoff, we suggest to use the $\mu^+/\mu^-$ ratio as a measurement
of the antiproton fraction in the cosmic rays below the GZK
cutoff.

In fact, the muons in the air showers are mainly from decays of
pions and kaons produced in the interactions of the primary cosmic
rays with the atmosphere \cite{Adair,muonratio}. The very high
energy secondary pion and kaon cosmic rays can be considered as
from the current fragmentation of partons in deep inelastic
scattering of the primary cosmic rays with the nucleon targets of
the atmosphere in a first approximation \cite{MSSY}. We also
consider only the favored fragmentation processes, i.e., the
$\pi^+$, which is composed of valence $u$ and $\bar{d}$ quarks, is
from the fragmentation of $u$ and $\bar{d}$ quarks in the nucleon
beam, and the $\pi^-$, which is composed of valence $\bar{u}$ and
$d$ quarks, is from the fragmentation of $\bar{u}$ and $d$ quarks
\cite{MSY}. Similarly, the $K^+$, which is composed of valence $u$
and $\bar{s}$, is from the fragmentation of $u$ and $\bar{s}$
quarks, and the $K^-$, which is composed of valence $\bar{u}$ and
$s$, is from the fragmentation of $\bar{u}$ and $s$ quarks. The
$\mu^+$ is from the decay of a $\pi^+$ or a $K^+$ and the $\mu^-$
is from the decay of a $\pi^-$ or a $K^-$. We can roughly estimate
the muon charge ratio by
\begin{equation}
\frac{\mu^+}{\mu^-} =\frac{\int_0^1 {\mathrm d} x \left\{\left[
u(x)+\bar{d}(x)\right]+\kappa
\left[u(x)+\bar{s}(x)\right]\right\}}{\int_0^1 {\mathrm d} x
\left\{\left[d(x)+\bar{u}(x)\right]+\kappa
\left[\bar{u}(x)+s(x)\right]\right\}}, \label{QME}
\end{equation}
where $q(x)$ is the quark distribution with flavor $q$ for the
incident hadron beam and $\kappa \sim 0.1 \to 0.3$ is a factor
reflecting the relative muon flux and fragmentation behavior of
$K/\pi$. Secondary collisions do not influence the above
estimation but can significantly reduce the final detected muon
energies, since the current parton beams still keep their flavor
content and act as the current partons after the strong
interactions with the partons in the atmosphere targets. Adopting
a simple model estimation of the parton flavor content in the
nucleon without any parameter \cite{ZZM}, we find that
$\mu^+/\mu^- \sim 1.7$ for proton and $\mu^+/\mu^- \sim 0.7$ for
neutron. This simple evaluation is in agreement with the empirical
expectation of $\mu^+/\mu^- \approx 1.66$ for proton and
$\mu^+/\mu^- \approx 0.695$ for neutron \cite{Adair} as well as
that in an extensive Monte Carlo calculation \cite{neutron-muon},
thus it provides a clear picture to understand the dominant
features for the muon charge ratio by the primary hadronic cosmic
rays. For the $\mu^+/\mu^-$ ratio for antiproton, it is equivalent
to the $\mu^-/\mu^+$ ratio for proton by using eq.~(\ref{QME}),
thus we find $\mu^+/\mu^- \sim 0.6$ for antiproton, which is close
to that for neutron. The $\mu^+/\mu^-$ ratio for antineutron is
also equivalent to the $\mu^-/\mu^+$ ratio for neutron, and it is
$\mu^+/\mu^- \sim 1.4$, which is close to that for proton. It is
hard to distinguish between the primary neutrons and antiprotons
(or protons and antineutrons) by the $\mu^+/\mu^-$ ratio of the
air shower, unless very high precision measurement is performed
and also our knowledge of the muon charge ratio for each nucleon
species is well established.

For the nucleon cosmic rays with energy above the pion
photoproduction, there is an admixture of neutrons and
antinucleons with the protons. It is hard to make a clear
distinction between the neutron and antiproton (proton and
antineutron) by the muon charge ratio $\mu^+/\mu^-$. Also the
number of nucleons should be very limited by the GZK cutoff
suppression. But there will be only protons and antiprotons in the
cosmic rays below the GZK cutoff from our above discussion. With
the collection of a sizeable number of events with enough
statistics, we expect to detect the antiproton content in the
cosmic rays by the measurement of the $\mu^+/\mu^-$ ratio. The
difference in the muon charge ratios between protons and
antiprotons will be enhanced if there is an energy cut on the
detected muons, as the ratio
$\left[u(x)+\bar{d}(x)\right]/\left[d(x)+\bar{u}(x)\right]$ (i.e.,
the dominant contribution in eq.~(\ref{QME})) has a strong $x$
dependence to increase for proton and decrease for antiproton at
large $x$. There could be also a composition of heavy nuclei in
the ultrahigh energy cosmic rays \cite{nuclei}. The nucleus beams
are estimated, by using eq.~(\ref{QME}), to have the muon charge
ratio $\mu^+/\mu^- \approx 1.1$, which is different from both of
these for protons and antiprotons. However, methods have been
developed to identify the primary nucleon and nuclei by other
information such as \v Cerenkov radiation \cite{Rawlins}. The
secondary muon size distributions have been also used in available
experiments to identify the mass composition of primary cosmic ray
extensive air showers \cite{Akeno,KASCADE}. We suggest to add muon
charge as a further information to identify the antiproton
composition. In combination with other information, it is thus
possible to detect the hadronic composition by measuring the muon
charge ratio for the cosmic ray events with ultrahigh energy.

We need to point it out here that none of the available facilities
for ultrahigh energy cosmic rays is aimed at measuring the muon
charge ratio as their physical motivation, so it is not practical
to expect the measurement of the muon charge ratio for the
ultrahigh energy comic ray showers within a short period. In
principle, there should be no difficulty to measure the muon
charge ratio by the available techniques of detectors
\cite{Kre99}, but this feature needs to be incorporated into
design of new generation facilities for the physical goal we
suggested. It is also not practical to measure the muon charge
ratio on an event by event basis. The reason is that not all
particles in an ultrahigh energy cosmic shower can be detected:
the detectors are distributed in arrays, and can not cover all
region. So some muons should be lost in the detection procedures.
Therefore the muon charge ratio should be studied from a large
number of events in a sense of statistics. Also for our goal of
identifying the nucleon species by their muon charge ratio, the
muons should be detected at distances near the cores of the cosmic
ray showers, and the energy of the muons should be high enough to
guarantee the detected muons of being from the cascading
fragmentation of the current partons in the primary cosmic rays in
consequent multi-scattering processes. We should point out that
there are technical limitations of the highest energy of muons to
be detected, so we should not expect that the detected muons are
from the the first fragmentation of the partons in the primary
cosmic rays, but rather the current partons in the primary cosmic
rays should still keep at the leading current partons in the
subsequent scattering processes so that their flavor information
is not completely lost at the highest energy of muons to be
detected~\footnote{Or the muon detector should be placed deep
underground for the purpose to detect the muons with initially
very high energy}.
 We should also point out that the flavor structure of the
partons is not explicit included in the available Monte Carlo
simulations, as the physics picture of the current fragmentation
we described above has not been incorporated in the available
Monte Carlo codes or models. In similar to the available air
shower profiles with muon size distributions \cite{Akeno},
detailed features with also muon charge and energy distributions
for the air shower profiles of primary proton and antiproton
cosmic rays are needed and waiting to be constructed by further
explicit researches. This can be only realized by new generation
experiments aiming at detecting the antiproton content in the
cosmic rays, by collections of a large number of events of cosmic
ray showers, together with their detailed muon charge and energy
information (perhaps with also orientation distribution in the
earth's magnetic field) contained in the measurements.

Our results not only work for the Z-burst hypothesis, they also
apply to any model involving the neutron and antiproton contents
in the cosmic rays. The cosmic rays above and the GZK cutoff can
be attributed as originated from the decays of relic metastable
superheavy particles clustered as dark matter in the galactic halo
\cite{X}, or from the topological defects that are left over from
the early-universe phase transitions caused by the spontaneous
breaking of symmetries \cite{Bha98}. There should be also some
composition of neutrons and antinucleons in the cosmic rays in
these models. In fact, the role played by the time dilation effect
of the neutrons has been discussed in a number of literature
\cite{Weiler,neutron}. But the combination with the proton/neutron
changing properties in their propagation makes this dilation
effect more significant. We thus obtain an interesting scenario
that the proton and neutron with energy above the GZK cutoff
change between each other upon interaction with the relic photons
in their propagation, and this indicates a non-trivial neutron
fraction in the nucleon cosmic rays above the GZK cutoff. Our
prediction of the propagation of neutrons and the suggestion of
using the muon charge ratio to detect the antiproton fraction of
the cosmic rays below the GZK cutoff, are applicable to all these
models.

In summary, we studied the role of neutrons in the cosmic rays by
taking into account the time dilation effect and the
proton/neutron changing properties in the pion photoproduction by
the relic photons. We predict a non-trivial neutron fraction in
the cosmic rays above the GZK cutoff, no matter the nucleon cosmic
rays come from sources with distance comparable to the GZK zone or
within local galaxies. However, there is a negligibly small
neutron fraction in the cosmic rays with energy below the GZK
cutoff, unless the sources are with distance comparable to 1 Mpc.
Thus, we also suggest to use the muon charge ratio to detect the
antiproton content of the cosmic rays below the GZK cutoff. This
may serve to provide some more information concerning several
existing models of using Z-bursts, superheavy particles, and
topological defects as origins for the observed events of cosmic
rays with energy above and near the GZK cutoff.
\\~


{\bf Acknowledgments: } This work is partially supported  by the
Taiwan CosPA Project funded by the Ministry of Education
(89-N-FA01-1-0 up to 89-N-FA01-1-5). It is also supported by the
National Natural Science Foundation of China under Grant Number
10421003 and by the Key Grant Project of Chinese Ministry of
Education (NO.~305001).





\end{document}